%% file: manuscript_arXivStyle_without_comments.tex
\documentclass[pra,aps,groupedaddress,superscriptaddress,notitlepage]{revtex4-1}

\input{./meta/packages}

\usepackage{amsthm,amsmath,amssymb}
\RequirePackage{hyperref}
\InputIfFileExists{gitinfo-latex.inc}{}{}

\usepackage{textcomp}
\usepackage{braket}
\usepackage{nicefrac}
\usepackage{units}
\usepackage{graphicx}
\begin{document}

\title{Neutral Impurities in a Bose-Einstein Condensate for Simulation of the Fröhlich-Polaron}
	\input{./meta/authors}

	\pacs{...}	
	
	\date{\today}
	
	\begin{abstract} %
We present an experimental system to study the Bose polaron by immersion of single, well-controllable neutral Cs impurities into a Rb Bose-Einstein condensate (BEC).
We show that, by proper optical traps, independent control over impurity and BEC allows for precision relative positioning of the two sub-systems as well as for independent read-out.
We furthermore estimate that measuring the polaron binding energy of Fröhlich-type Bose polarons in the low and intermediate coupling regime is feasible with our experimental constraints and limitations discussed.
\end{abstract}

	\maketitle

\section{Introduction}
The immersion of single, controllable atoms into a Bose-Einstein condensate (BEC) realizes a paradigm of quantum physics - individual quantum objects interacting coherently with a single or few mode bath.
This system allows to experimentally address various questions of quantum engineering, including local, non-demolition measurement of a quantum many-body system \cite{NgBose2008}; cooling of qubits while preserving internal state coherence \cite{Daley2004, Griessner2006}; or engineering bath-mediated, long-range interaction between two or more impurities \cite{Klein2005}.
An impurity strongly interacting with the quantum gas will loose its single particle properties and it is rather described in terms of quasi particles, which are known as Bose polarons \cite{Cucchietti2006, Tempere2009, Grusdt2014}. 
Particularly for strong interaction, such systems have been predicted to show remarkable properties such as self-trapping \cite{Cucchietti2006} or polaron clustering \cite{Klein2007}. 
Experimentally, impurities in BECs have been introduced, for example, as many atoms of different internal state \cite{Palzer2009, Fukuhara2013} or different atomic species \cite{Ospelkaus2006, Scelle2013}, or as individual ions \cite{Zipkes2010, Schmid2010} or electrons \cite{Balewski2013}. 

The system we report on here considers a single, neutral  impurity in an ultracold gas, where the effect of impurity-impurity interactions, either direct or mediated by the bath, can be neglected. The corresponding Hamiltonian thus reflects an extremely imbalanced mixture and can be written as \cite{Tempere2009, Shashi2014}
\begin{eqnarray}
\hat{H} &=& \sum_\mathbf{k} \varepsilon_\mathbf{k} \hat{a}_\mathbf{k}^\dagger \hat{a}_\mathbf{k}^{\vphantom{\dagger}} + \frac{1}{2} \sum_{\mathbf{k},\mathbf{k}^\prime, \mathbf{q}} V_{BB}(\mathbf{q})\, \hat{a}_{\mathbf{k}^\prime - \mathbf{q}}^\dagger \hat{a}_{\mathbf{k}+\mathbf{q}}^\dagger \hat{a}_\mathbf{k}^{\vphantom{\dagger}} \hat{a}_{\mathbf{k}^\prime}^{\vphantom{\dagger}} \nonumber \\
 & & + \frac{\hat{p_I}^2}{2 m_I} + \sum_{\mathbf{k}, \mathbf{q}} V_{IB}(\mathbf{q})\, \hat{\rho}_I(\mathbf{q})\, \hat{a}_{\mathbf{k}-\mathbf{q}}^\dagger \hat{a}_\mathbf{k}^{\vphantom{\dagger}}.\label{eq:BosePolaron}
\end{eqnarray}
The first row of Eq.~(\ref{eq:BosePolaron}) represents the BEC part of the Hamiltonian, where $a_\mathbf{k}^\dagger$ ($a_\mathbf{k}^{\vphantom{\dagger}}$) creates (annihilates) a boson of mass $m_B$, momentum $\mathbf{k}$, and energy dispersion $\varepsilon_\mathbf{k} = \hbar^2 k^2 /(2 m_B)$. 
The interaction of bosons within the BEC is given by the contact $s$-wave interaction with Fourier transform $V_{BB}(\mathbf{q})$.
The impurity atom of momentum $\hat{p}_I$,  mass $m_I$ and density $\hat{\rho}_I$ as well as its interaction with the BEC via the potential $V_{IB}$ are described by the second row in Eq.~(\ref{eq:BosePolaron}). 

\subsection{The Fröhlich polaron}
Originally, the polaron concept was developed for condensed matter systems to describe electrons moving in a crystal lattice. 
The interaction between a moving electron and lattice of the ion cores forms a propagating quasi-particle called polaron, comprising electron and surrounding phonon cloud. 
Specifically the effective mass of the polaron as well as its energy could strongly differ from the bare electron's values depending on the interaction with the crystal \cite{Landau1933, Pekar1951}. 
In the limit of very small electronic wave vector, the underlying crystal structure can be neglected, and the crystal can be described as a continuously polarizable medium.
In this case, the model to describe polarons is the well-known Fr\"ohlich Hamiltonian ~\cite{Froehlich1954} 
\begin{eqnarray}\label{eq:FrohlichPolaron}
\hat{H}_{p} = \frac{\hat{p_I}^2}{2m_I} + \sum_{\mathbf{k} \neq 0} \hbar \omega_\mathbf{k} \hat{b}_\mathbf{k}^\dagger \hat{b}_\mathbf{k} + \sum_{\mathbf{k} \neq 0} V_\mathbf{k} e^{i\mathbf{k \cdot \hat{r}}} \left( \hat{b}_\mathbf{k} + \hat{b}_{-\mathbf{k}}^\dagger \right),
\end{eqnarray}
given by the sum of the kinetic energy of the impurity particle with mass $m_I$, the energy of the phonons in the medium with dispersion $\omega_\mathbf{k}$, and the interaction energy with coupling constant $V_\mathbf{k}$ arising between the two of them. The coupling strength is essentially determined by the dielectric constant of the crystal and can be quantified by a single, dimensionless coupling constant $\alpha$. 
The regime of small $\alpha$ is characterized by weak and intermediate polaron coupling while for values of $\alpha \gg 1$ the strong coupling regime is realized. Simple perturbation theory predicts the crossover around a value of $\alpha \approx 6$.
Experimentally,  Fr\"ohlich polarons have been observed in condensed matter systems for small and moderate interaction strengths $\alpha$, however, the regime of so-called strong coupled polarons with $\alpha\gg1$ has so far not been experimentally investigated \cite{Hodby1969, Landolt2002}.

Recently, the quantum gas Bose-polaron was subject of intense theoretical work.
It was shown that the description of a Bose-polaron  Eq.~(\ref{eq:BosePolaron})  can be directly mapped onto the Fröhlich Hamiltonian  Eq.~(\ref{eq:FrohlichPolaron}) via a Bogoliubov transform~\cite{Tempere2009} and that this model could be accessed experimentally~\cite{Grusdt2014,Shashi2014}.
This analogy holds as long as the interaction between two BEC excitations can be neglected.
The resulting Hamiltonian has the same operator structure as Eq.~(\ref{eq:FrohlichPolaron}), where the crystal (optical) phonons are replaced by elementary Bogoliubov excitations of the BEC with dispersion $\varepsilon_\mathbf{q} = \hbar v_s q \sqrt{1 + (\xi\,q)^2 /2}$, where $v_s = \hbar/(\sqrt{2} m\,\xi)$ is the sound velocity in the BEC, $\xi = 1/\sqrt{8 \pi n_\mathrm{BEC}\,a_{BB}}$ is the healing length, $n_\mathrm{BEC}$ is the BEC atomic density, and $a_{BB}$ is the $s$-wave scattering length between two atoms in the BEC.
The polaronic coupling parameter in the dilute quantum gas scenario ~\cite{Tempere2009,Grusdt2014} 
\begin{equation}\label{eq:CouplingStrength}
\alpha = \frac{a_\text{IB}^2}{a_\text{BB}\xi}
\end{equation}
is given by the $s$-wave scattering lengths between a Boson of the quantum gas with the impurity ($a_\text{IB}$) or with other quantum gas atoms ($a_\text{BB}$), and the condensate healing length $\xi$. 
For the quantum gas Bose-polaron, these parameters can be varied over a large range of values in the weakly interacting regime and the strong coupling regime by means of Feshbach resonances.
Moreover, the characteristic properties of the emerging polaron which are the binding energy $E_\text{p}$ and its effective mass $m_\text{p}$ can be inferred by, e.g., radio frequency spectroscopy and trap frequency measurements, respectively~\cite{Shashi2014}.
Importantly, also the Bose polaron beyond the Fröhlich model can be experimentally realized yielding access to a rich and highly controllable model system of impurity physics in quantum fluids.

\section{Experimental Realization}
Our experimental approach to realizing the Bose polaron aims at immersing single or few neutral Caesium ($^{133}$Cs) atoms into a Rubidium ($^{87}$Rb) BEC.
This combination of species features several advantages, facilitating the realization, control and characterization of the Bose polaron.
First, due to the relatively high nuclear charge of Rb and Cs the fine structure splitting of both species is also relatively large and allows to tune dipole traps in between the two fine structure lines of the first excited $P$-level with moderate unwanted photon scattering \cite{LeBlanc2007, Arora2012}.
As a consequence the atoms of this element do not experience a dipole potential.
In order to improve the control over both species independently, we employ this fact constructing a species-selective lattice allowing for trapping and controlled transport of impurity atoms, only.

Second, for a dipole trap wavelength of $\lambda = 1064\,$nm, the trapping frequency $\omega$ and thus the gravitational sag $g/\omega^2$ with $g$ the gravitational acceleration, is equal to the percent level for the two species.

Third, using few or even single impurity atoms has several advantages. 
For Cs representing the minority component, three-body losses limiting the lifetime of the polaron are due to Rb-Rb-Cs collisions rather than Cs-Cs-Rb collision, where the loss coefficient $L_3$ of the former is an order of magnitude smaller than for the latter \cite{Spethmann2012}.
While three-body losses will still be a limitation of the atomic lifetime, the lifetime of the polaron is expected to be significantly larger than the lifetime in balanced Rb-Cs mixtures. However, if the elastic interaction of multiple impurities is desired this can be controlled by adjusting the number of trapped impurities.
Furthermore, fluorescence imaging of such a small number of atoms in an optical lattice allows for single-site resolved detection of the impurities with standard optical systems \cite{Karski2009}.
Moreover, only the dynamics of a single impurity allows to dynamically track the trajectory of a polaron, because crossing of the trajectories of multiple indistinguishable impurities would lead to an ambiguity.

For the design of the experimental apparatus, additional considerations have to be made. In a combined system of quantum gas and single atoms, the respective ways to experimentally extract information from averages differ: For a quantum gas, a single realization yields an ensemble average of typically $10^3 \ldots 10^5$ atoms. For single atoms, in contrast, averages have to be formed as time averages of typically $10^2 \ldots 10^3$  repetitions for identical parameter values. For a combination, the statistics is clearly limited by probing the single impurity, while usually the time scale of a single experimental run is limited by the production of a BEC. 

In the following we first discuss the experimentally relevant parameter ranges and constraints for our system of single Cs impurities in a Rb BEC, before we turn to the presentation of our experimental apparatus.

\begin{figure}[h!]
	\includegraphics{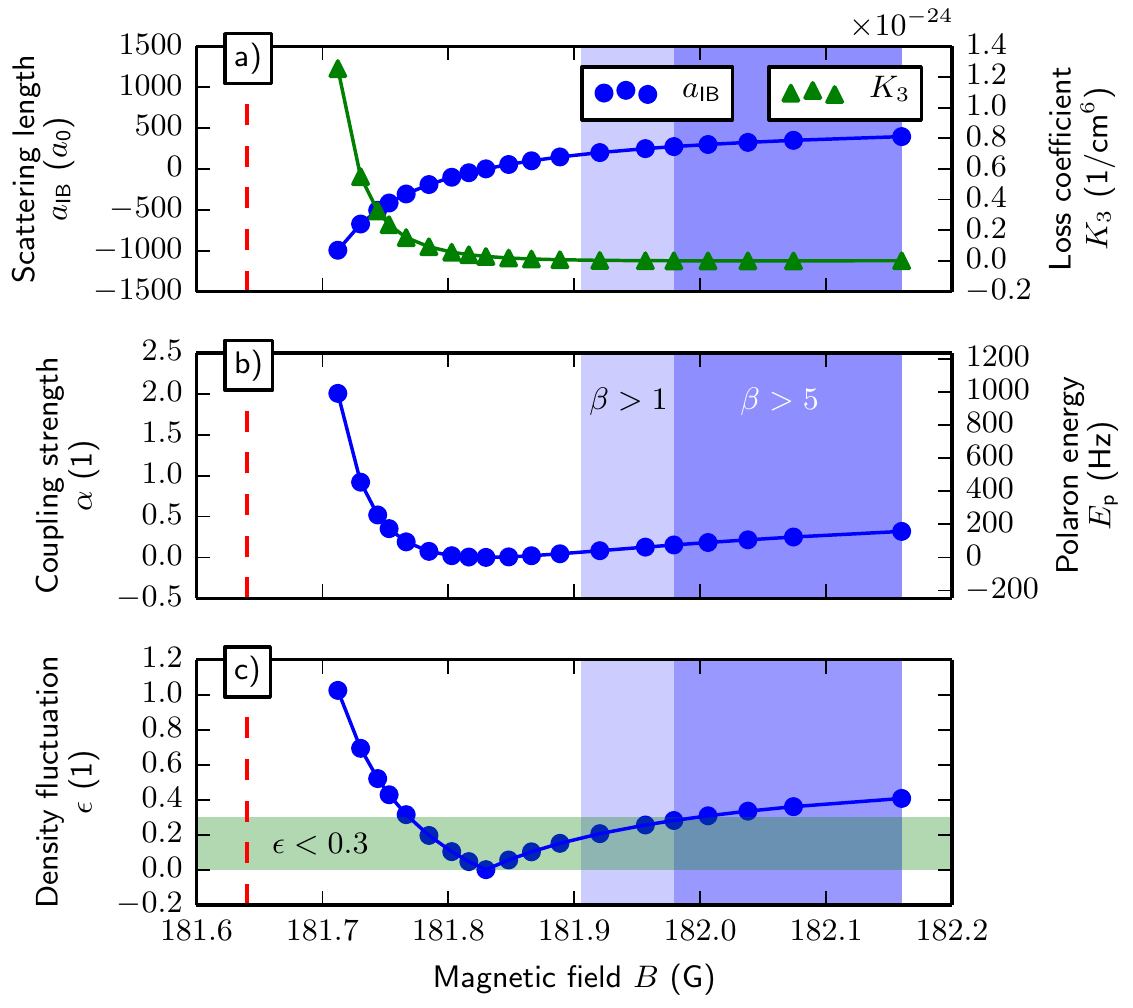}
	\label{figure:PolaronProperites}
	\caption{\textbf{Bose polaron properties at the Feshbach resonance.}\\
		a) shows the theoretical values of the interspecies scattering length $a_\text{IB}$ and the three-body loss coefficient $K_3$ for Rb-Rb-Cs scattering at the \unit[181]{G} Feshbach resonance (dashed red).
		For our typical BEC density of $10^{14} \,\text{cm}^{-3}$ the estimated coupling strength $\alpha$ and the corresponding polaron energy $E_\text{p}$ is shown in b).
		The vertical blue shaded area indicates fulfillment of the condition $\beta = \nu_\text{p}/\gamma_\text{p} > 5$ ($>1$ light shaded).
		c) The parameter $\epsilon$ is derived in \cite{Grusdt2014} and gives an estimate of how well the Bose polaron fits the Fröhlich type.
		The horizontal green shaded area indicates $\epsilon<0.3$ which means good correspondence to the Fröhlich type polaron.}

\end{figure}

\subsection*{The Rb-Cs Bose-Polaron}
For the experimental characterization of the Bose-Polaron, we focus on the binding energy $E_\text{p} = h \nu_\text{p}$ in the following.
This energy can be measured by radio or microwave spectroscopy, driving a Zeeman or hyperfine transition between two impurity states, where one state is interacting with the bosonic bath forming the polaron, whereas the other state is non-interacting \cite{Schirotzek2009, Kohstall2012}.
The decay rate of the polaron $\gamma_\text{p}$ implies a lower bound for the linewidth of the polaron spectroscopy peak and is dominated by three-body losses, i.e.~molecule formation, occurring with rate $\gamma_\text{p} = n_\text{BEC}^2 L_3$ with the loss coefficient $L_3$ and the BEC density $n_\text{BEC}$.
In order to clearly resolve the polaron peak at a frequency shift of $\nu_\text{p}$, the ratio $\beta = \nu_\text{p}/\gamma_\text{p}$ should be significantly larger than 1 yielding a figure of merit for the determination of optimum experimental parameters.

In Figure~\ref{figure:PolaronProperites} we explore the regimes of Bose polarons realized with our typical experimental parameters discussed below, employing theoretical data of elastic scattering length $a_\text{IB}$ and three-body loss coefficient $L_3$ around the interspecies Feshbach resonance at \unit[181]{G}~\cite{Wang2012}. 
The range of polaronic coupling strengths $\alpha$ accessible with our experimental setup is determined by several parameters: the boson-boson scattering length $a_\text{BB}\approx 100 a_0$ \cite{vanKempen2002}, with the Bohr radius $a_0 = \unit[53]{pm}$, is given by the background scattering length of Rb and does not change within the range of magnetic field values considered here; BEC peak densities are on the order of $n_\text{BEC,max} = 10^{14}\,\text{cm}^{-3}$ which serves as a worst case approximation for $n_\text{BEC}$; and the interspecies scattering length $a_\text{IB}\approx 650\,a_0$ at zero magnetic field is tunable in the vicinity of the Feshbach resonance \cite{Pilch2009, Takekoshi2012}. 

From this we calculate the polaron coupling constant $\alpha$ and polaronic binding energy $E_\text{p}$ using a simple analytic expression for an impurity with infinite mass from \cite{Shashi2014}, see Figure~\ref{figure:PolaronProperites}(b). Furthermore, the polaronic decay rate $\gamma_\text{p}$ is calculated from the BEC peak density and theoretical values for $L_3$ in vicinity of the Feshbach resonance~\cite{Wang2012}. 
For $\beta > 1$, we can reliably resolve polarons spectroscopically within the discussed parameter ranges, indicated by the vertical shaded regions in Figure~\ref{figure:PolaronProperites}.

Furthermore, in Figure~\ref{figure:PolaronProperites}(c) we use the parameter $\epsilon$ from \cite{Grusdt2014} to indicate to which extend the Bose polaron realized can be described by means of the Fröhlich model discussed above, where $\epsilon \ll 1$ corresponds to a good description. 
One can see that $\epsilon$ lies well below 1 for the experimentally directly accessible range (blue). For reference, the value of $\epsilon<0.3$ which is given in \cite{Grusdt2014} is marked as horizontally shaded area in Figure~\ref{figure:PolaronProperites}. 

For observation of a strongly coupled Fröhlich polaron, clearly, the parameter range has to be adjusted by, for example, employing Feshbach resonances with more favorable three-body loss properties or optical tuning methods \cite{Compagno2014}, by increasing the sensitivity of the spectroscopic detection for a given life-time, and by adapting the BEC density via changes of the trap geometry.
However, a Bose-polaron simulating the Fröhlich model in the weak and intermediate coupling regime is well accessible by immersing single neutral Cs atoms into a BEC with density of the order of $\unit[10^{14}]{cm^{-3}}$.
 
\subsection*{Rb Bose-Einstein Condensate}
In order to optimize the statistics of single atom probing, the experimental setup aims at a short BEC production time. 
This is realized by, first, a short initial laser cooling stage, where a 3D magneto-optical trap (MOT) is loaded from a 2D MOT in  $\approx \unit[2.5]{s}$ and, second, evaporation in a steep optical dipole trap which is formed by a horizontal and a vertical beam within $\approx \unit[4]{s}$.
In order to avoid perturbation from cooling and trapping of single Cs atoms in a MOT, the Rb cloud is prepared in the magnetic field insensitive $F=1, m_F =0$ state while evaporating.
We typically prepare a BEC with $2.5\times10^4$ atoms at a peak density of $\unit[1.2 \times 10^{14}]{cm^{-3}}$ and a critical temperature of approximately \unit[120]{nK}. The BEC's decay rate of \unit[0.4]{Hz} is dominated by two- and three-body collisions at this density but decreases to \unit[0.08]{Hz} for a lower number of atoms. %
For technical details of the BEC production and state preparation, see Section \ref{sec:Methods}.

\subsection*{Single Atoms}
Single or few Cs atoms are captured in a high magnetic field gradient ($\approx 250\,$G/cm) MOT \cite{Haubrich1996}, spatially overlapped with the Rb MOT, but operated at a different time in the preparation sequence. 
In contrast to the Rb MOT, the Cs MOT is loading atoms from the background, and the laser cooling beams have a smaller diameter and beam intensity, leading to an overall reduced loading rate.
\begin{figure}
\includegraphics{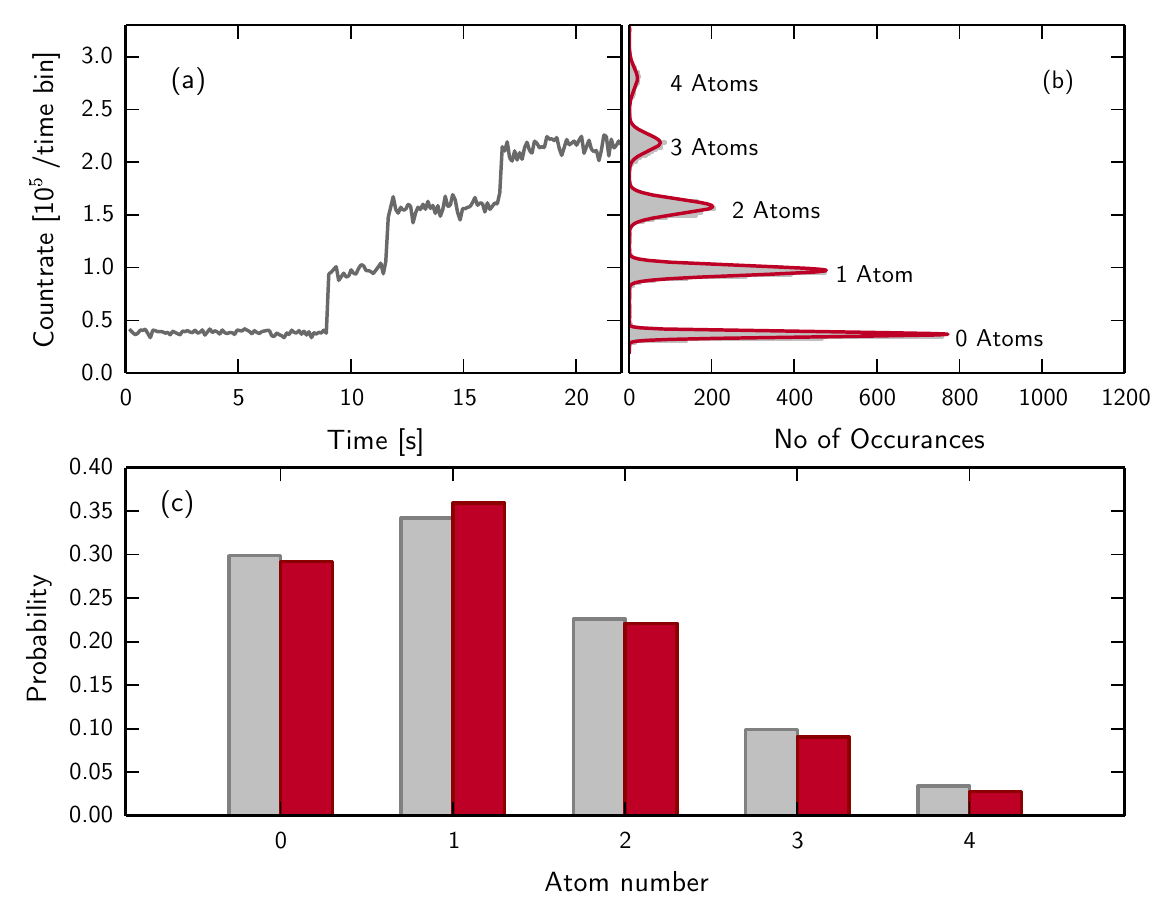}
\label{fig:loadingcurve}
\caption{\textbf{Loading curve and histogram of the high gradient MOT.} (a) Typical loading trace of the single atom MOT. 
	After \unit[8]{s} a first atom is loaded into the MOT and in steps of \unit[3-5]{s} an atom is gained.
	The time is binned in 110 ms to collect enough fluorescence photons.
	(b) Histogram of the electron count rate taken over 30 traces with 267 data points each. The red solid line shows a Multi-Gaussian fit to the data.
	(c) Probabilities for each atom number obtained from the area of each peak in the multi-Gaussian fit (gray bars) and from a Poisson distribution (red bars) with mean atom number n=1.23, calculated from the histogram.
}
\end{figure}
The fluorescence signal of such a MOT features discrete values, which can be assigned to a specific number of atoms in the MOT. 
A Poissonian distribution describes the loading statistics in the few atom regime, observed in a histogram of measured fluorescence signal, corresponding to the distribution of atom numbers in Figure~\ref{fig:loadingcurve}~(b). 
The width of each histogram peak is ideally given by the shot noise of the atoms' fluorescence light; practically, however, fluctuations of the MOT laser beam intensities and technical contributions such as readout noise additionally broaden the peaks. 
This limits the maximum countable atom number to roughly eight for our setup. 
Refer to Section \ref{sec:Methods} for technical details and a description of the fluorescence imaging system.

In order to provide position control over the single Cs atoms independently from the Rb BEC trap, we apply a species selective optical conveyor belt lattice, formed by two counter propagating, linearly polarized laser beams with wavelength $\lambda_\mathrm{lat} = \unit[790.0180]{nm}$ and a waist of $\unit[29] {\mu m}$.
For this wavelength between the Rb D-Lines, the resulting potential cancels out for Rb atoms \cite{LeBlanc2007, Arora2012}, but at the same time the frequency is blue detuned for Cs, providing tight confinement along the lattice axis in the nodes of the standing wave with depths up to $7300\,E_r^\mathrm{Cs} = \unit[850]{\mu K}\times \mathrm{k_B}$, with $E_r^\mathrm{Cs} = \hbar^2 k^2 /2 m_\mathrm{Cs}$ the single photon recoil energy, $k=2 \pi/\lambda_\mathrm{lat}$,  $\mathrm{k_B}$ the Boltzmann constant, and $m_\mathrm{Cs}$ the mass of a Cs atom.
While the lattice provides tight axial confinement for Cs atoms, it does not confine the atoms radially.
We therefore superpose the lattice axis with one beam of the dipole trap (see Section~\ref{sec:Methods}) and obtain a maximum trap depth of \unit[1.45]{mK} radially,
resulting in trap frequencies of $\unit[2\pi \times 4]{kHz}$ radially and $\unit[2\pi \times 460]{kHz}$ axially. 
The lifetime of atoms in the lattice at full depth is limited to $\tau \approx \unit[1.24]{s}$ by phase fluctuations.
If in addition an optical molasses is used to cool the atoms, the lifetime can be extended up to $\tau = \unit[71]{s}$, limited by background pressure.

An important quantity characterizing the lattice is the selectivity $s = E_r^\mathrm{Rb}/E_r^\mathrm{Cs}$ for a given intensity and wavelength.
By performing Raman-Nath \cite{Martin1988, Cahn1997, Gadway2009} scattering on the BEC for various lattice wavelengths, we have identified the optimal wavelength $\lambda_\mathrm{lat,vac} = (790.0180 \pm 0.0017)\,$nm  and find a selectivity of $5000:1$ for Cs.
In order to transport the Cs atoms by a defined distance, a precisely controlled relative detuning $\delta$ between the lattice beams is used, which causes the standing wave interference pattern to move at a velocity $v=\lambda_\textrm{lat} \delta /2$ for a specific amount of time \cite{Kuhr2003B}. For details also see Section~\ref{sec:Methods}.

\subsection*{Combining Single Atoms with the Quantum Gas}

During evaporation, the Rb cloud is prepared in the magnetic field insensitive state $\ket{F=1,m_F=0}$ by a radio frequency transition, see Section~\ref{sec:Methods}. 
After it has been sufficiently cooled down to be well localized in the dipole trap crossing region, the Cs MOT is switched on.
\begin{figure}
	\includegraphics{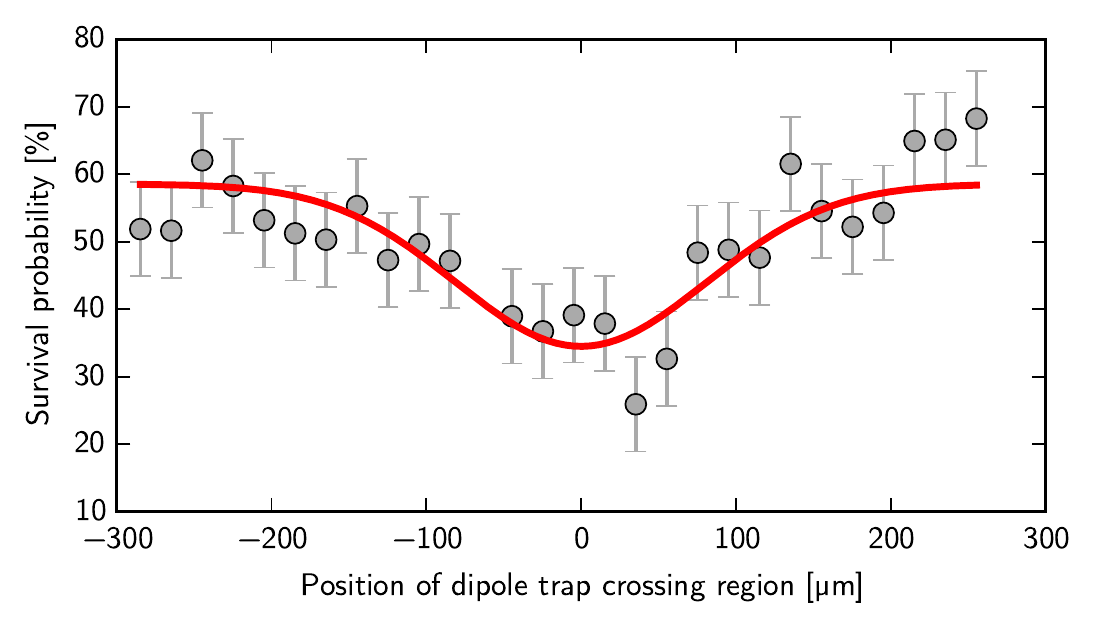}
	\label{figure:lightAssistedCollisions}
	\caption{\textbf{Light assisted collisions.} The graph shows the probability of Cs atoms surviving a \unit[1.3]{s} MOT period while a cold Rb cloud is present in the dipole trap crossing region in dependence of the crossing region's position along the horizontal dipole trap beam. We attribute the losses to light assisted collisions of Cs with Rb in the presence of near resonant Cs MOT light. A negative Gaussian fit (red) serves as a guide to the eye.
	}
\end{figure}
To avoid immediate losses of the single Cs atoms due to light-induced collisions \cite{Spethmann2012InsertingSingleCs,Weber2010,Weiner1999}, both species are trapped in different traps with a displacement of $\unit[110]{\mu m}$.
Figure~\ref{figure:lightAssistedCollisions} shows the probability of Cs atoms surviving a \unit[1.3]{s} MOT period while a cold thermal Rb cloud is present in the dipole trap crossing region in dependence of the crossing region's position.
The minimum in survival probability marks the position of maximum overlap between MOT and dipole trap.

As an initial step towards immersing the Cs atoms into the Rb BEC, we immerse them into a thermal Rb gas.
\begin{figure}
	\includegraphics{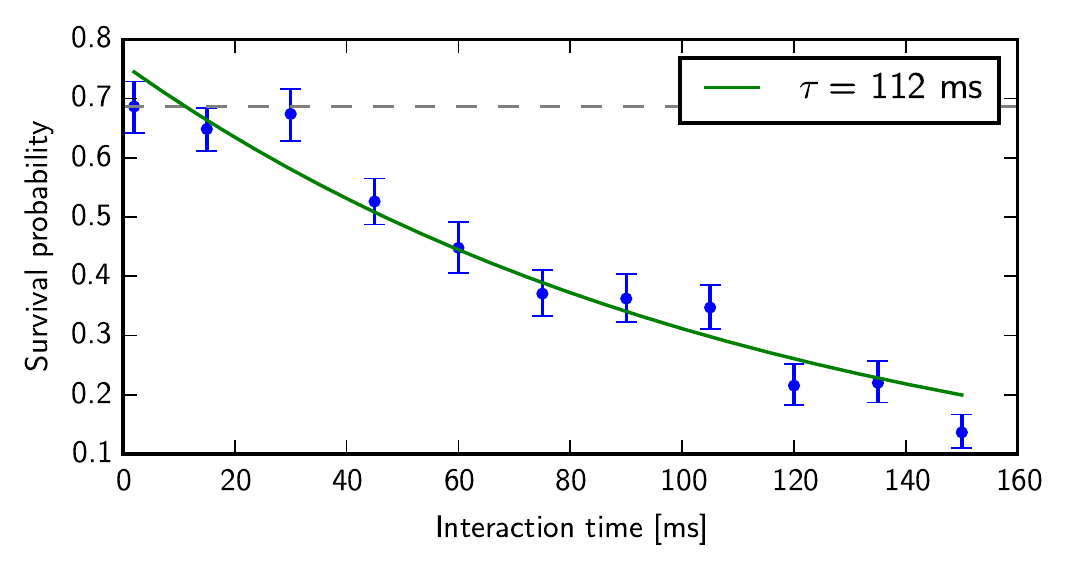}
	\label{figure:CsLifetime}
	\caption{\textbf{Lifetime of Cs in presence of a cold Rb cloud.} The graph shows the probability of recapturing Cs atoms from the dipole trap after they have been stored together with a cold Rb cloud for a given 'interaction time'.
		Without Rb present in the trap, the survival probability is 0.69 (dashed gray). For a Rb cloud in $\ket{F=1,m_F=0}$ with $38000$~atoms at a temperature of $T=\unit[3]{\mu K}$ at trap frequencies of $\omega = 2 \pi \times \left(63,1300,1300\right) \unit{Hz}$ and a peak density of $\rho_0 = \unit[1.3 \times 10^{13}]{cm^{-3}}$, the fit reveals a lifetime of $\tau = \unit[112]{ms}$ for Cs atoms in $\ket{F=3}$.
	}
\end{figure}	
Here, the horizontal dipole trap is sufficiently deep to directly trap Cs atoms from the MOT.
After switching off the Cs MOT, we let the two species interact for a certain time before pushing Rb out of the trap by shining in resonant light perpendicular to both dipole trap beams.
The high gradient MOT is switched on to recapture the Cs atoms from the dipole trap and the number of atoms in both MOT phases is compared to each other.
Figure~\ref{figure:CsLifetime} shows a measurement of the survival probability in dependence of the time between switching off the Cs MOT and pushing out the Rb, which we refer to as 'interaction time'. For a Rb cloud in $\ket{F=1,m_F=0}$ with $\unit[38\times10^3]{atoms}$ at a temperature of $T=\unit[3]{\mu K}$ at trap frequencies of $\omega = 2 \pi \times \left(63,1300,1300\right) \unit{Hz}$ and a peak density of $\rho_0 = \unit[1.3 \times 10^{13}]{cm^{-3}}$, we measure a lifetime of $\tau = \unit[112]{ms}$ for Cs atoms in $\ket{F=3}$.
This is lower than expected because with a 3-body loss coefficient of $L_3 = \unit[5\times 10^{-26}]{cm^6 s^{-1}}$ \cite{Spethmann2012} the 3-body loss-rate can be calculated to be $\nicefrac{1}{\tau_{\mathrm{3-body}}} = L_3 \times \rho_0^2 =\nicefrac{1}{\unit[116]{ms}}$  at peak Rb density $\rho_0$ and the Cs atoms should see a  lower average density.
We have not yet fully examined which processes limit the lifetime, but we expect two-body losses to play a role since the two species are not yet optically pumped to their lowest Zeeman substate.

\section{Methods}\label{sec:Methods}
\begin{figure}
	\includegraphics{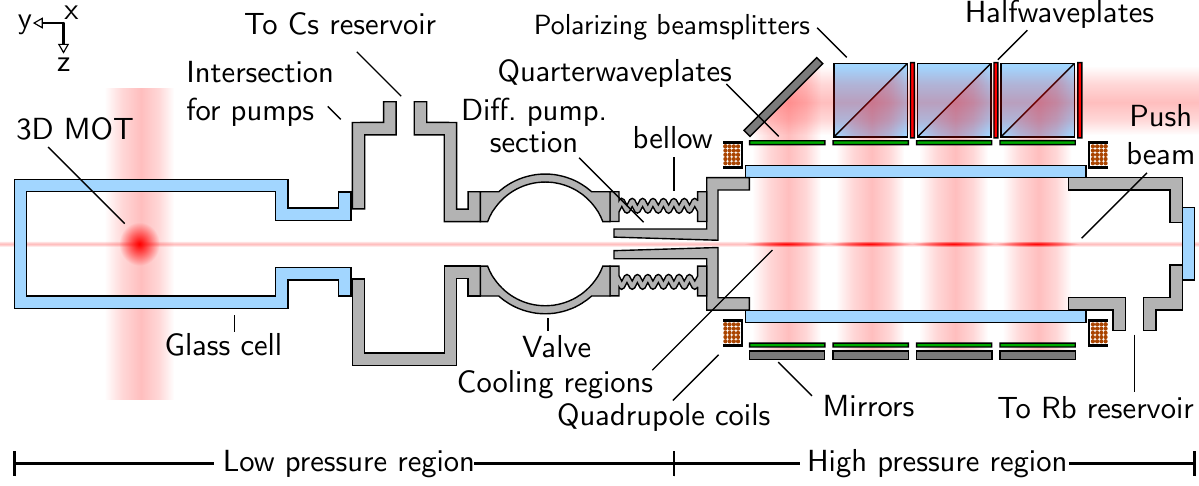}
	\label{figure:MotSchematics}
	\caption{\textbf{Schematical top view of our 2D/3D MOT system.}
	}
\end{figure}

\begin{figure}
	\includegraphics{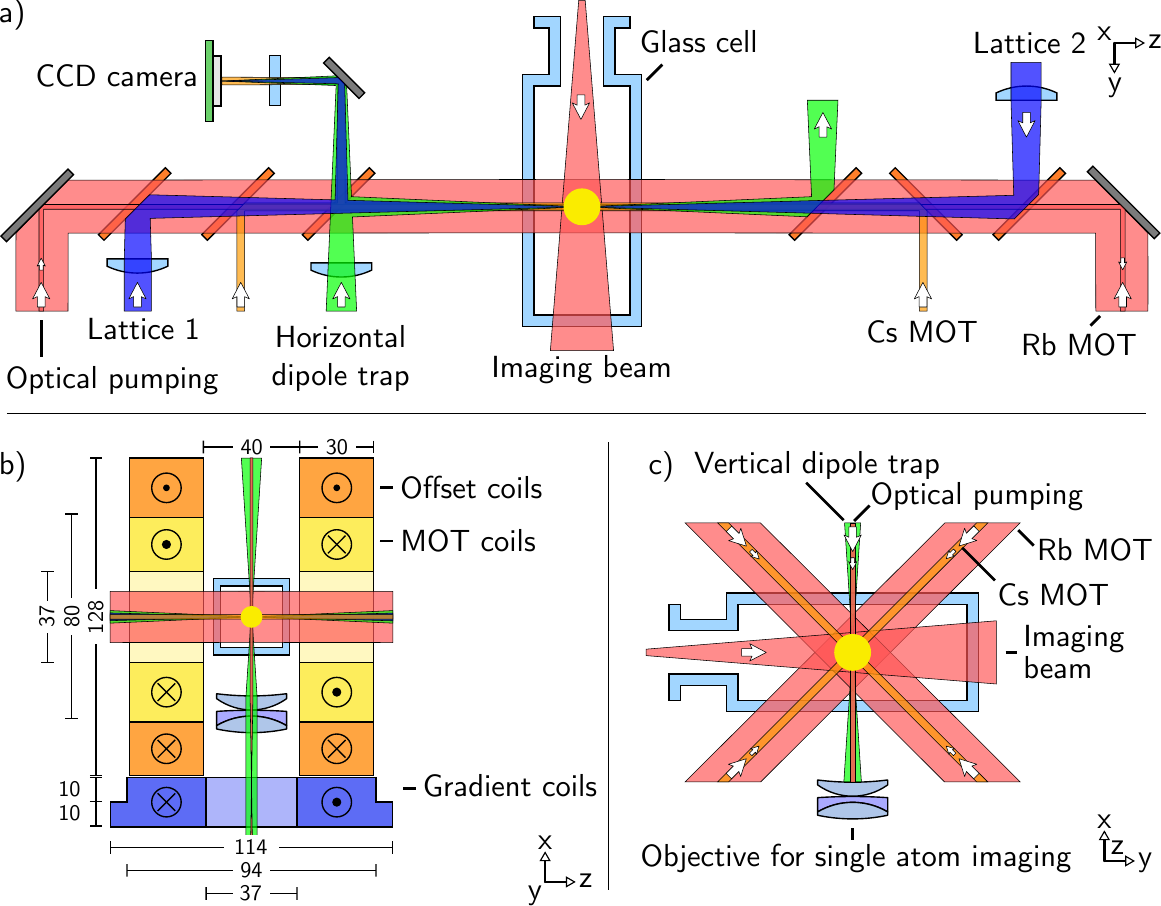}
	\label{figure:expSideOverview}
	\caption{\textbf{Overview of the experimental setup.} a) Experiment in top view: With dichroic mirrors beams for the optical dipole trap, MOTs for each species, the optical lattice for Cs, and an optical pumping beam for Rb are overlapped. A CCD camera resembles the position of the cold atoms which is useful for the initial alignment of the dipole trap and the optical lattice to the position of the Cs MOT. b) Front view with main magnetic coils (dimensions in mm). c)~Side view.
	}
\end{figure}

Our experiments take place in a two-chamber vacuum system (see Figure~\ref{figure:MotSchematics} and \ref{figure:expSideOverview}) consisting of a low pressure ($\approx 10^{-10}$~mbar) and a high pressure region ($\approx 10^{-7}$~mbar), separated by a differential pumping section of length \unit[83]{mm} and diameter increasing from \unit[1.8]{mm} to \unit[4]{mm}.
The low pressure region is formed by a glass cell, whereas the high pressure region is located in a titanium chamber and contains the 2D MOT for $^{87}$Rb which is loading atoms from the background gas. The distance between the 3D MOT region and the 2D MOT is approximately $30$~cm.

The coil system shown in Figure~\ref{figure:expSideOverview}~b) is installed at the low pressure side, providing a quadrupole field for the Cs and Rb MOTs, a homogeneous field to address Feshbach resonances, and a vertical magnetic field gradient for Stern-Gerlach experiments.
Additionally, compensation coils are installed to provide weak, homogeneous fields in all three dimensions up to $\unit[2]{G}$  to compensate magnetic stray fields.

\subsection*{Cooling and Trapping}

\paragraph*{Magneto Optical Traps for $^{87}$Rb}

The 2D MOT \cite{Dieckmann1998} setup is based on a design described in \cite{Gericke2007a,Utfeld2006}.
The elongated titanium vacuum chamber features four windows with a clear aperture of 110~mm x 25~mm as an optical access for the MOT beams.
Four distinct cooling regions are formed by orthogonal retro-reflected beams with diameters of $2 \times w_0 = \unit[25]{mm}$, and a viewport allows for an additional near-resonant beam ('push-beam') along the cooling axis to increase the flux of the atomic beam.
It is 1.9~MHz blue detuned from the $F=2 \rightarrow F^\prime = 3$ transition with an intensity of \unitfrac[120]{nW}{cm$^2$}.
The windows are surrounded by two quadrupole coil pairs, each coil consisting of 130 windings of 1~mm enameled copper wire.
We typically operate the MOT at magnetic field gradients of $\unitfrac[11]{G}{cm}$ in vertical and horizontal direction. %
Our 3D MOT features a standard six beam configuration, where detuning and intensity can be controlled independently of the 2D MOT beams.
\begin{figure}
	\includegraphics{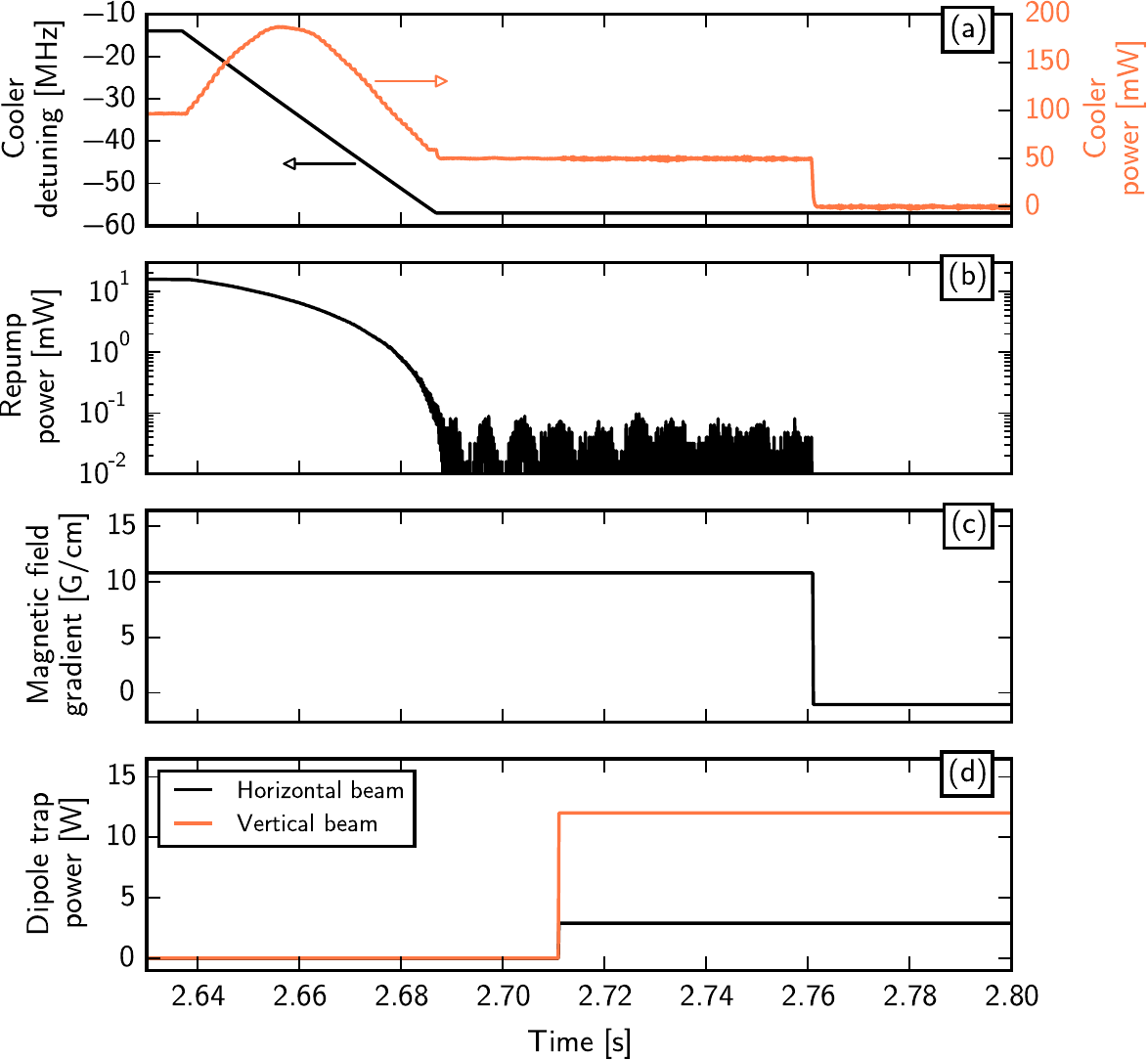}
	\label{figure:CMOT}
	\caption{\textbf{CMOT scheme.} We achieve an optimal transfer rate of Rb from the MOT to the dipole trap by increasing the red-detuning of the cooler while reducing the power of the repumper.
	}
\end{figure}
We usually load the 3D MOT from the 2D MOT atomic beam for $\unit[2.5]{s}$ at a loading rate of $\unitfrac[10^9]{atoms}{s}$ before we enter a compressed MOT ('CMOT') phase \cite{Ketterle1993,Lewandowski2003} (see  Figure~\ref{figure:CMOT}). 
The parameters for these Rb MOTs are listed in Table~\ref{table:motparams}.
\begin{table}
	\begin{tabular}{cccc}
		\hline  \hline
		& 2D MOT  & 3D MOT & CMOT  \\ 
		\hline
		cooler power			& 240~mW    		& 100~mW		 & 50~mW\\
		cooler detuning   		& -13.2~MHz  & -13.9~MHz  & -58~MHz\\
		repumper power			& 28~mW	 		&30~mW		 &14.8~\textmu W\\
		repumper detuning       & -5~MHz 	& -5~MHz & -5~MHz\\
		beam size $w_\mathrm{0}$& \unit[12.5]{mm}	&	\unit[10]{mm}	 & \unit[10]{mm}		\\
		magnetic field gradient & $~\unitfrac[11]{\mathrm{G}}{\mathrm{cm}}$ & \multicolumn{2}{c}{$\unitfrac[6]{G}{cm}$ radially}\\ 
		&  &\multicolumn{2}{c}{$\unitfrac[11]{G}{cm}$ axially}   \\

	\end{tabular}
	\caption{Parameters for the operation of different MOTs for $^{87}$Rb.}
	\label{table:motparams}
\end{table}

\begin{figure}
	\includegraphics{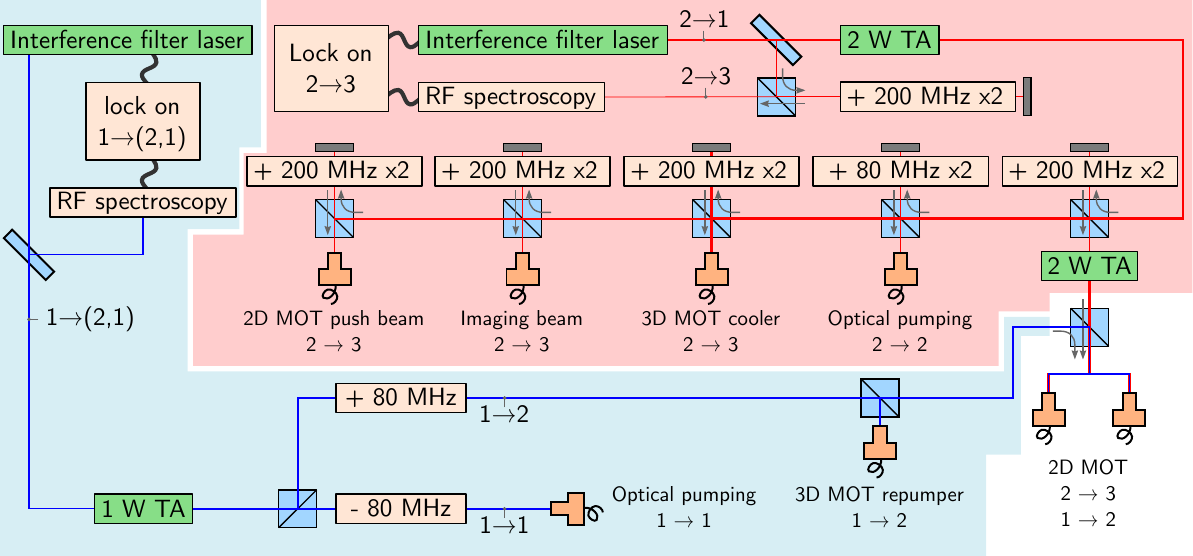}
	\label{figure:RubidiumLasersystem}
	\caption{\textbf{Schematic diagram of our laser setup for Rubidium trapping, manipulation and detection.} The notation $F \rightarrow F'$ refers to the hyperfine transition of the $D_2$ line to which the light is resonant, while $F \rightarrow (F'_{1},F'_{2})$ refers to a crossover signal of the Doppler-free spectroscopy.}
\end{figure}

The laser light for operation of Rubidium MOTs, internal state manipulation and detection is provided by the master oscillator power amplifier (MOPA) setup shown in Figure~\ref{figure:RubidiumLasersystem}.
We employ interference-filter-stabilized extended-cavity diode lasers \cite{Baillard2006} (Radiant Dyes NarrowDiode), frequency stabilized to a frequency-modulation spectroscopy of Rubidium 87.
The lasers' output power of $\approx \unit[50]{mW}$ is amplified by home-made tapered amplifiers (TAs) up to \unit[2]{W}, frequency shifted by acousto-optical modulators (AOMs) and transported to the experiment via polarization maintaining, optical single-mode fibers (Nufern PM780-HP).

\paragraph*{Single Atom MOT}
The position of the Cs MOT is overlapped with the Rb MOT as both use the same coil system. 
A set of 6 diaphragms with variable aperture mounted in the coil holders is used to align the MOT beams to precisely overlap in the middle of the glass cell. 
The cooling light is \unit[10]{MHz} red detuned to the F=4~$\rightarrow$~F'=5 transition of the Cs D2 line and has a total power of typically $\unit[500]{\mu W}$. 
The repumping light is on resonance to the F=3~$\rightarrow$~F'=3 transition with a total power of typically $\unit[15]{\mu W}$. 
In every beam pair a piezo driven mirror at \unit[110]{Hz} frequency destroys phase coherence between orthogonal beam pairs and therefore avoids interference effects, which lead to an instable MOT position. 
To keep the MOT loading time short, a low magnetic field gradient of \unitfrac[40]{G}{cm} axially and \unitfrac[20]{G}{cm} radially is applied for \unit[20]{ms}.
During this sufficiently short time the trap volume is large enough to load on average one atom \cite{Kuhr2003}. 
In a next step the magnetic field gradient is increased in $\unit[8]{ms}$ to \unitfrac[275]{G}{cm} in axial and \unitfrac[140]{G}{cm} in radial direction. 
This effectively pins the atom number and avoids additional atom loading during the imaging process.
The duration of the low gradient stage depends on the vapor pressure of the Cs atoms and the required atom number.
For an efficient transfer of the trapped atoms into the optical dipole trap, a low MOT temperature is needed. Therefore we increase the red detuning of the cooler light to \unit[72]{MHz} for \unit[50]{ms}, while setting its power so low that we just do not lose the atoms. We release the atoms into the dipole trap by switching off the MOT beams. The cooler is switched off $\unit[2]{ms}$ after the repumper ensuring that the atoms remain in their lowest fine-structure state $\ket{F=3}$.

\paragraph*{Optical Dipole Trap}
Our optical dipole trap is a crossed beam trap at 1064~nm. The trap is formed by a horizontal beam with a focal waist of $w_0 = \unit[22]{\mu m}$ at $\unit[4]{W}$ and a vertical beam with a focal waist of $w_0 = \unit[165]{\mu m}$ at $\unit[12]{W}$ of power.
We measure trap frequencies of $\unit[3.7]{kHz}$ radial to the horizontal beam and $\unit[120]{Hz}$ along the horizontal beam, while the axial confinement is mainly caused by the vertical beam.
The beam setup allows both for a forced evaporation scheme \cite{Clement2009} as well as standard, passive evaporation \cite{Arnold2011}.
We keep the pointing instability minimal by using optical fibers to transport the light to the experiment (LMA-PM-15 by NKT Photonics for the vertical beam and Liekki Passive-10/125-PM for the horizontal beam).
AOMs between the laser light source (Nufern NUA-1064-PD-0050-D0 fiber amplifier) and the optical fibers are controlled by a PID controller that determines and stabilizes the laser power at the experiment with a bandwidth of $\unit[110]{kHz}$.

\begin{figure}
	\includegraphics{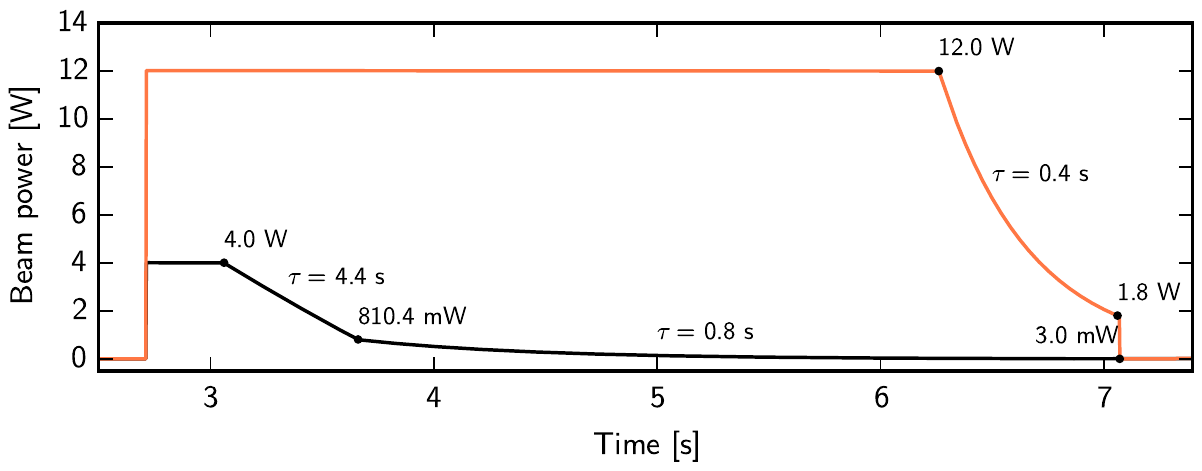}
	\label{figure:evap}
	\caption{\textbf{Evaporation scheme.} The graph shows the horizontal beam's power (black) and the vertical beam's power (orange) over time, including parameters of the exponential ramps.
	}
\end{figure}

\paragraph*{Evaporative Cooling}
Both beams are switched on at full power during the CMOT phase.
At the end of the CMOT phase, the MOT lasers and the quadrupole field are switched off.
The repumping light is switched off $\unit[200]{\mu s}$ before the cooling light, so that the atoms are pumped to the $F=1$ state and approx. $10^6$ atoms are transferred to the dipole trap.
Figure~\ref{figure:evap} shows our subsequent evaporation scheme.
After $\unit[300]{ms}$ of self-evaporation we decrease the horizontal beam's power exponentially in two sections with different decay constants $\tau$.
As soon as the increase in density within the crossing region leads to three-body losses, we also ramp down the power of the vertical beam until a BEC with $2.5\times 10^4$~atoms forms after a total evaporation time of $\unit[4.3]{s}$. The entire BEC sequence, including 3D MOT loading, CMOT phase and evaporation ramps has been optimized with an evolutionary algorithm, that is implemented as a part of our timing software \cite{EvolAlgo}.

\paragraph*{Internal State Preparation}
The Rb atoms are prepared in the magnetically insensitive state $\ket{F=1,m_F=0}$ during evaporation so that the magnetic field gradient of the single atom MOT does neither heat nor destroy the cold Rb cloud:
During self-evaporation, a magnetic field of $\unit[1.4]{G}$ is switched on to lift the degeneracy of the Zeeman substates.
Then the atoms are pumped to the $\ket{F=1,m_F=1}$ state by a $\unit[900]{\mu s}$ light pulse resonant to the $F=1 \rightarrow F'=1$ $\sigma^+$-transition and a $\unit[500]{\mu s}$ light pulse resonant to the $F=2 \rightarrow F'=2$ $\pi$-transition of the $D_2$ line. 
Both beams are turned on simultaneously.
Their respective powers are $\unit[300]{nW}$ and $\unit[480]{\mu W}$ with red detunings of $\unit[2.6]{MHz}$ and $\unit[1]{MHz}$ at equal beam waists of $\unit[1.2]{mm}$.

\begin{figure}
	\includegraphics{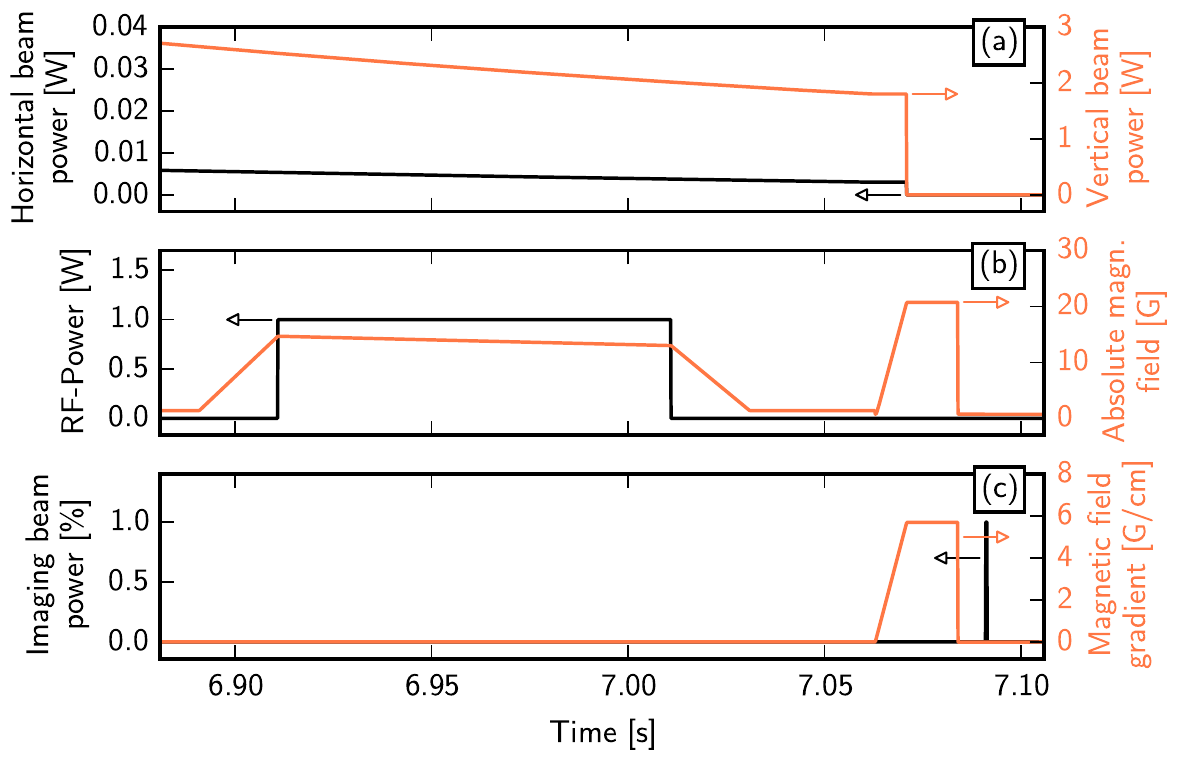}
	\label{figure:landauZener}
	\caption{\textbf{Driving a Landau-Zener transition and performing a Stern-Gerlach experiment.}
		Rb atoms in the $\ket{F=1, m_F=1}$ state are transfered to the $\ket{F=1, m_F=0}$ state by performing a Landau-Zener sweep.
		The magnetic field is ramped such that the detuning of the radio frequency with respect to the $\ket{m_F=1} \rightarrow \ket{m_F=0}$ transition changes from far blue detuning to far red detuning.
		After the atoms are released from the dipole trap, a magnetic field gradient is applied in order to perform a Stern-Gerlach Experiment.
	}
\end{figure}

By applying a magnetic field gradient of typically $\unitfrac[5.7]{G}{cm}$ during a $\unit[13]{ms}$ time of flight experiment (see Fig.~\ref{figure:landauZener}), we perform a Stern-Gerlach experiment to measure the population of magnetic substates.
Without optical pumping we observe an almost equally distributed spin mixture, while with optical pumping approximately $\unit[90]{\%}$ of the atoms are in the $m_F=1$ Zeeman substate and approximately $\unit[10]{\%}$ remain in $m_F=0$. Optical pumping does not deplete the number of atoms in the condensate.

After optical pumping we transfer the population to the $m_F=0$ state with a Landau-Zener sweep as shown in Figure~\ref{figure:landauZener}:
We apply a radio frequency at $\unit[10.176]{MHz}$ for $\unit[100]{ms}$ while we increase the magnetic field linearly.
The magnetic field ramp is chosen such that the detuning of the radio frequency with respect to the $\ket{m_F=1} \rightarrow \ket{m_F=0}$ transition falls from approximately $+ \unit[550]{kHz}$ to $- \unit[550]{kHz}$.
We observe a  Rabi-frequency of $\Omega = 2\pi \times \unit[2.1]{kHz}$ with a transfer efficiency of $\unit[95]{\%}$, with no negative impact on the number of atoms in the BEC.

\subsection*{Imaging}
\begin{figure}
	\includegraphics{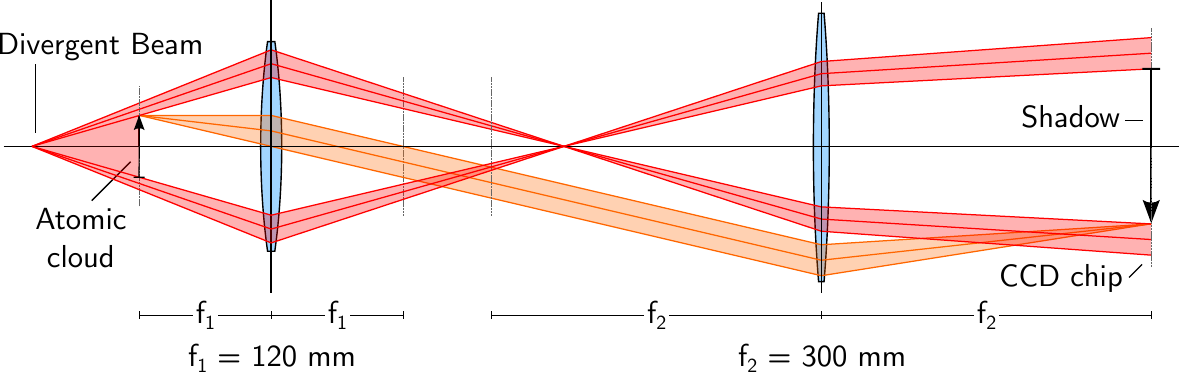}
	\label{figure:ImagingRaytrace}
	\caption{\textbf{Absorption imaging with a divergent beam.}
		The drawing shows both virtual rays originating from the atomic cloud and real rays from the divergent imaging beam. The atomic cloud lies in the focal plane of the imaging system. Independent of where the divergent beam's focus is located, a shadow magnified by $\nicefrac{f_2}{f_1}$is cast onto the CCD chip.
	}
\end{figure}

We image the Rb cloud by absorption imaging with linearly polarized light resonant to the $\ket{F=1}\rightarrow \ket{F'=1}$ transition of the $D_2$ line, see Figures~\ref{figure:expSideOverview} and \ref{figure:ImagingRaytrace}.

To observe the single atoms we use fluorescence imaging. 
Near resonant light from the MOT beams is scattered in the whole solid angle. We collect 3.3\% of the photons corresponding to $\unitfrac[0.012]{pW}{atom}$ for a saturation parameter $S_0=I/I_s=1$ by a custom made objective.
It has a numerical aperture (NA) of 0.36 and is placed beneath the glass cell at a distance of \unit[30.3]{mm} to the atoms' position (see Figure~\ref{figure:expSideOverview}c). 
Stray light protection is crucial and the whole beam path is located inside black anodized lens tubes and mirror housings. 

The objective tube is mounted on a precision xyz linear stage (Newport Ultralign Model 562-XYZ) for accurate positioning.
The light which is collimated by the objective is focused onto an EMCCD camera (Andor iXon 3 897) by a lens with focal length $f=\unit[1000]{mm}$.
This yields a magnification of M=33. The camera chip has $\unit[512 \times 512]{pixel}$ with a pixel size of $\unit[16 \times 16]{\mu m}$ which leads to a field of view of $\unit[250]{\mu m}$. 
An advantage of this setup is that the objective can be aligned independently from the focusing lens which is producing the image on the camera. 
\begin{figure}
	\includegraphics{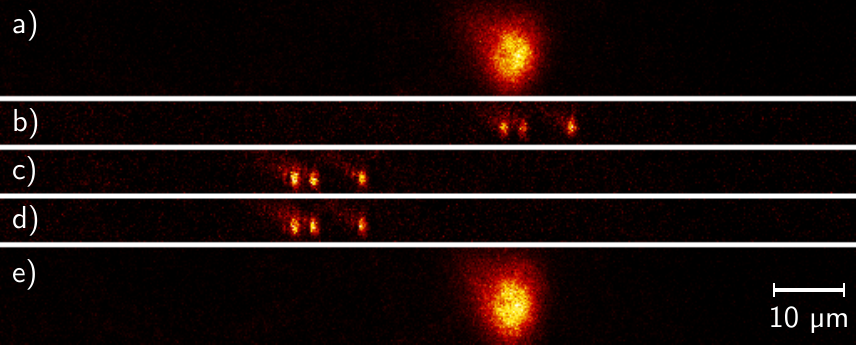}
	\label{fig:Transport}
	\caption{\textbf{Fluorescence images of single Cs atoms in MOT and lattice.} Four consecutive images of three single Cs atoms. a) Fluorescence image of Cs atoms trapped in a MOT, b) after being transferred to the species-selective conveyor belt lattice, c) after being moved for 30 µm, d) after being held in place, e) after being transferred back to the MOT.
	}
\end{figure}
The high EMCCD gain and the quantum efficiency of nearly 60\% of the camera allows to observe single atoms in the MOT as well as in the optical lattice (see Figure~\ref{fig:Transport}).
During imaging the atoms are illuminated by the repumper light of the MOT and cooler light, which is \unit[20]{MHz} red detuned in order to compensate the light shift caused by the dipole trap.
Exposure times of \unit[150]{ms} (\unit[400]{ms}) in the MOT (lattice) are usually used to image the atoms. The number of atoms in the MOT can be determined with close to 100~\% reliability from the brightness of the image.

\subsection*{Optical Lattice}
\begin{figure}
\includegraphics[scale=0.5]{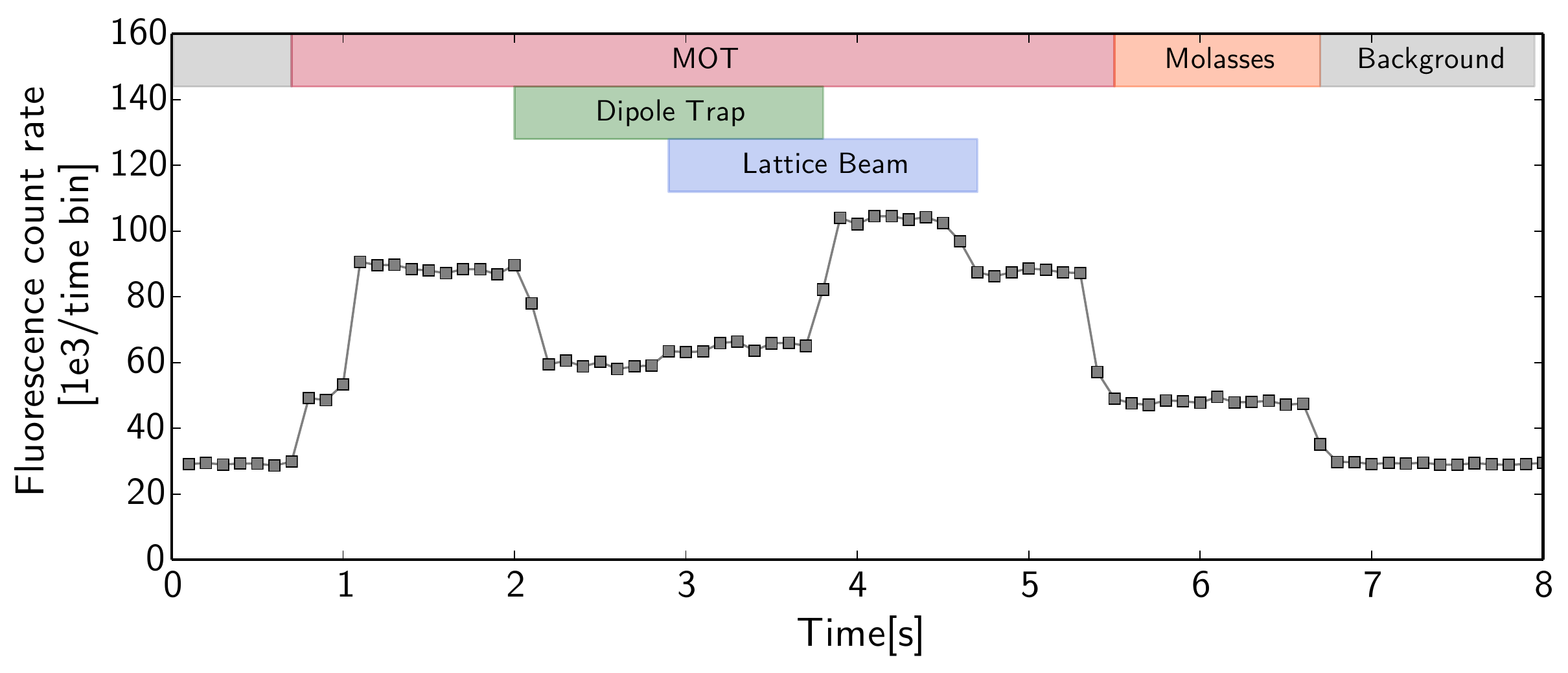}
\label{fig:LightShift}
\caption{\textbf{AC Stark effect of dipole trap and lattice.} A typical fluorescence trace with lattice and dipole trap beam. During the MOT phase one atom is captured. Switching on the dipole trap reduces the fluorescence level by about 70\%. In combination with the lattice a slight increase is observed because it compensates the light shift of the dipole trap. For the combination of MOT and lattice an increase of about 10\% in fluorescence level is observed. 
	In the end the count rate due to stray light and the dark count rate is detected. 
	A binning time of 100 ms is used.
}
\end{figure}

Our optical lattice consists of a two beam setup (see Figure~\ref{figure:expSideOverview}), where both beams are guided to the experiment by means of optical fibres with identical length.
This helps to reduce phase drifts and guarantees a good beam quality at the experimental side.
To make sure that both beams hit the small-volume Cs MOT, we exploit the AC stark effect.
The dipole trap is red detuned and hence reduces the fluorescence light of the MOT, whereas the lattice frequency is blue detuned and enhances the fluorescence (see Figure~\ref{fig:LightShift}).

\subsection*{Lattice Transport}

For the single atom transport in the conveyor belt optical lattice \cite{Kuhr2003B}, a relative detuning $\delta$ between the two lattice beams is employed to create a standing wave pattern, moving at velocity $v=\lambda_\textrm{lat} \delta /2$.
For a typical transport, the detuning between the beams is ramped in a trapezoid shape: a linear ramp from zero to the maximum detuning yields an acceleration of the atoms in the lattice, followed by a plateau of constant detuning where the atoms move at constant velocity.
After this plateau, the detuning is ramped back to zero and the atoms are decelerated again. 
The transport distance is given by the integral over the velocity. 
Thereby large transport distances in the range of millimeters can be realized, which are only limited by trap size.
This is in contrast to phase shifting transport approaches, where the maximum transport distance is given by the maximum phase shift \cite{Belmechri2013}.
The absolute maximum acceleration $a_\text{max} = k U_0 / m_\text{Cs} \approx 4\cdot10^{5}\,\text{m}/\text{s}^2$ in the lattice is determined by the Cs mass $m_\text{Cs}$, the laser wave number $k$ and the potential depth $U_0 \approx \unit[1.5]{mK}$ of the standing wave potential.

\begin{figure}
	\includegraphics[scale=0.8]{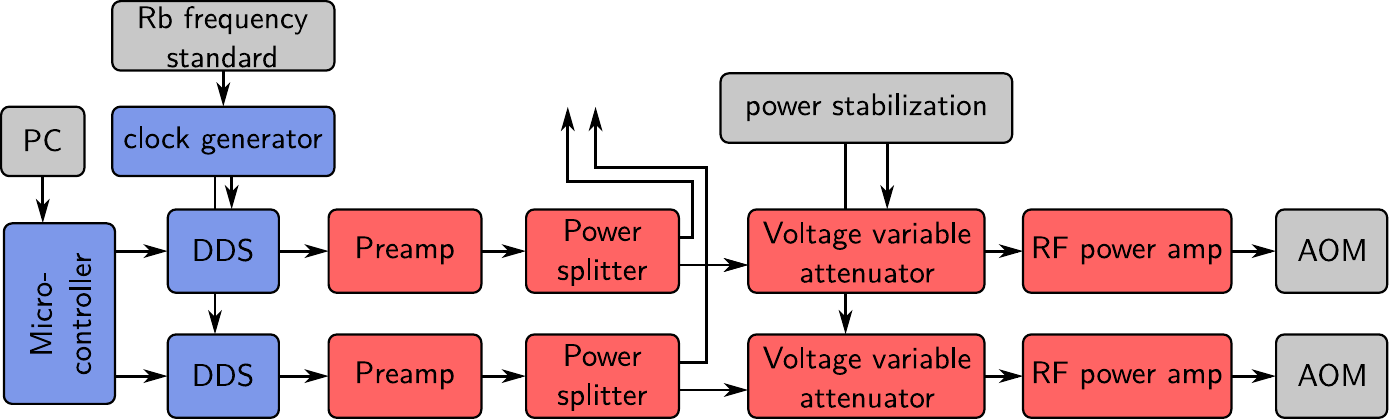}
	\label{figure:DDSSetup}
	\caption{\textbf{Lattice control setup.}
		The DDS-based signal generation (blue boxes) supplies two phase stable radio-frequencies to the amplifier chains (red boxes) which drive the lattice AOMs. 
	}
\end{figure}
In order to control the detuning, each lattice beam is frequency shifted by a common value of $f_1 \approx f_2 = \unit[160]{MHz}$ in an AOM double-pass setup. 
The relative detuning between the two beams then is given by $\delta = f_1 - f_2$ and can be controlled by the two RF frequencies supplied to the AOMs.
Both RF signals have to be phase-stable compared to each other and the offset-frequency of both channels has to be equal up to a fraction of a Hertz to yield a stable standing wave pattern when both beams interfere for $\delta = 0\,$Hz.
Therefore, we use a driver electronics based on direct digital synthesis ('DDS') with an amplifier chain shown in Figure~\ref{figure:DDSSetup}.
The  DDS chips employed (AD9954) sample the output sine wave in a digital circuit and have an analog digital converter stage for signal output.
By supplying both DDS chips with the same clock signal which is locked to a Rb frequency standard, the relative phase stability between the signals is guaranteed by the digital sampling of the output signal.
The frequency of the output sine wave is set by a \unit[32]{bit} control parameter, yielding a frequency resolution of \unit[0.09]{Hz} which allows for very small frequency variations compared to the output frequency of \unit[80]{MHz}.
To supply output powers of up to \unit[$\sim$30]{dBm} a two stage amplifier chain combined with a voltage controlled attenuator for output power control is applied. The RF power level can be directly used to adjust and stabilize the light intensity of the lattice beams. 
The DDS chips are controlled by a microcontroller which provides exact timing of frequency changes and stores the data for the frequency ramps.

\section{Conclusion}\label{sec:Conclusion}
Our experimental results constitute an ideal starting point for studying coherent interactions of single Cs impurities in a Rb BEC. 
We have shown that, for this system, the experimental realization, control and detection of a Bose polaron in the weak and intermediate coupling regime of the Fröhlich model is accessible. Moreover, our experimental apparatus not only facilitates the creation of a tunable Bose polaron but it also allows for controlled positioning and dynamical studies of impurities in a quantum fluid.

\section*{Acknowledgements}
The project was financially supported partially by the European Union via the ERC Starting Grant 278208 and partially by the DFG via SFB/TR49. 
D.M. acknowledges funding by the graduate school of excellence MAINZ, F.S. acknowledges funding by Studienstiftung des deutschen Volkes, and T.L. acknowledges funding from Carl-Zeiss Stiftung. 
\bibliographystyle{bmc-mathphys} %
\bibliography{bmc_article}      %

\end{document}

%% file: meta/packages.tex
\usepackage[utf8x]{inputenc}
\usepackage{color}
\usepackage{bbm} 
\usepackage{amsfonts,amsmath,amssymb,stmaryrd}
\usepackage{graphicx}
\usepackage{subfigure}  
\usepackage{bbm} 
\usepackage{hyperref}
\usepackage[pdftex]{epsfig}
\usepackage{mathrsfs}
\usepackage{verbatim}
\usepackage{centernot}
\usepackage{ulem}
\usepackage{array}
\usepackage{cancel}
\usepackage{ifthen}
\usepackage{bm}	
\usepackage{siunitx}
\usepackage{todonotes}
\InputIfFileExists{gitinfo-latex.inc}{}{}

%% file: meta/authors.tex
\author{Michael Hohmann}
\affiliation{Department of Physics and Research Center OPTIMAS, University of Kaiserslautern, Germany}

\author{Farina Kindermann}
\affiliation{Department of Physics and Research Center OPTIMAS, University of Kaiserslautern, Germany}

\author{Benjamin Gänger}
\affiliation{Department of Physics and Research Center OPTIMAS, University of Kaiserslautern, Germany}

\author{Tobias Lausch}
\affiliation{Department of Physics and Research Center OPTIMAS, University of Kaiserslautern, Germany}

\author{Daniel Mayer}
\affiliation{Department of Physics and Research Center OPTIMAS, University of Kaiserslautern, Germany}
\affiliation{Graduate School Materials Science in Mainz, Gottlieb-Daimler-Strasse 47, 67663 Kaiserslautern, Germany}

\author{Felix Schmidt}
\affiliation{Department of Physics and Research Center OPTIMAS, University of Kaiserslautern, Germany}
\affiliation{Graduate School Materials Science in Mainz, Gottlieb-Daimler-Strasse 47, 67663 Kaiserslautern, Germany}

\author{Artur Widera}
\affiliation{Department of Physics and Research Center OPTIMAS, University of Kaiserslautern, Germany}
\affiliation{Graduate School Materials Science in Mainz, Gottlieb-Daimler-Strasse 47, 67663 Kaiserslautern, Germany}